\documentclass[aip,prb,twocolumn,reprint,preprintnumbers,amsmath,amssymb,floatfix]{revtex4-1}

\usepackage[colorlinks,linkcolor=blue,citecolor=blue]{hyperref}
\usepackage{graphicx}
\usepackage{dcolumn}
\usepackage{bm}
\usepackage[subnum]{cases}
\usepackage{amsmath}
\usepackage{siunitx}
\usepackage{url}
\usepackage{xcolor}
\usepackage{comment}

\begin{document}

\title{Coherent transport properties of a three-terminal hybrid superconducting interferometer}
\author{F. Vischi}
\affiliation{Dipartimento di Fisica, Universit\`{a} di Fisica, I-56127 Pisa, Italy}
\author{M. Carrega}
\affiliation{NEST, Istituto Nanoscienze-CNR  and Scuola Normale Superiore, I-56127 Pisa, Italy}
\author{E. Strambini}
\affiliation{NEST, Istituto
Nanoscienze-CNR  and Scuola Normale Superiore, I-56127 Pisa, Italy}
\author{S. D'Ambrosio}
\affiliation{NEST, Istituto
Nanoscienze-CNR  and Scuola Normale Superiore, I-56127 Pisa, Italy}
\author{F.S. Bergeret}
\affiliation{Centro de Fisica de Materiales (CFM-MPC), Centro Mixto CSIC-UPV/EHU,
Manuel de Lardizabal 5, E-20018 San Sebasti´an, Spain}
\affiliation{Donostia International Physics Center (DIPC),
Manuel de Lardizabal 5, E-20018 San Sebasti´an, Spain}

\author{Yu. V. Nazarov}
\affiliation{Kavli Institute of Nanoscience, Delft University of Technology, Lorentzweg 1, 2628 CJ, Delft, The Netherlands}

\author{F. Giazotto}
\email{f.giazotto@sns.it}
\affiliation{NEST, Istituto
Nanoscienze-CNR  and Scuola Normale Superiore, I-56127 Pisa, Italy}
\begin{abstract}
We present an exhaustive theoretical analysis of  a double-loop Josephson proximity interferometer, as the one recently realized by Strambini \emph{et
al.} for the  control  of the Andreev spectrum via an external magnetic field. This system, called $\omega$-SQUIPT, consists of a T-shaped diffusive normal
metal (N)  attached to three superconductors (S) forming a double loop configuration.  By using the quasiclassical Green function formalism, we  calculate
the local normalized density of states, the Josephson currents through the device and the dependence of the former on the length  of the junction arms, the applied
magnetic field and the S/N interface transparencies. We show that by tuning the fluxes through the double loop, the system undergoes transitions from a
gapped to a gapless state. We  also evaluate the Josephson currents flowing in the different arms as a function of magnetic fluxes and explore the  quasi-particle
transport, by considering a metallic probe tunnel-coupled to the Josephson junction and calculating its I-V characteristics.
Finally, we study the performances of the $\omega$-SQUIPT and its potential applications, by  investigating its electrical and magnetometric properties.
\end{abstract}

\maketitle

\section{Introduction}
The superconducting quantum interference proximity transistor (SQUIPT)~\citep{giazotto_superconducting_2010} is a 
new concept of superconducting interferometer based on the proximity effect\cite{McMillan1968,usadel_generalized_1970} in a normal (N) metallic nanowire embedded in a superconducting (S) loop.
The phase-controlled density of states (DoS) of the proximized nanowire makes the SQUIPT an ideal building block for the realization of heat nanovalves~\citep{strambini_proximity_2014} or very sensitive and ultra-low power dissipation magnetometers~\citep{ronzani_highly_2014,dambrosio_normal_2015,meschke,virtanen2016} able to succeed the state-of-the-art SQUID technologies, with particular interest in the single-spin detection~\citep{giazotto_hybrid_2011}.  

\begin{figure}[t!]
\includegraphics[width=8.5cm]{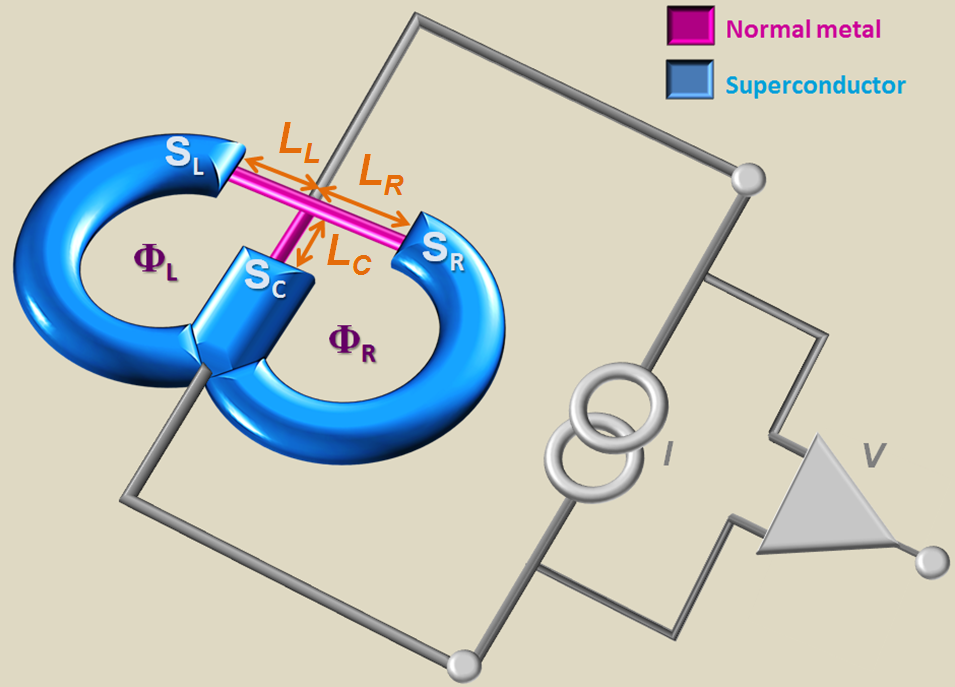}
\vspace{-0mm} 
\caption{
(Color online) Scheme of the $\omega$-SQUIPT in a current-biased setup. $I$ is the current flowing through the circuit, $V$ is the voltage drop across the device. $\Phi_L$ and $\Phi_R$ represent the magnetic fluxes piercing the left and right loop, respectively. $L_L$, $L_C$, and $L_R$ refer to the left, center and right arms length of the T-shaped normal metal, respectively. Finally, $S_L$, $S_C$, and $S_R$ refers to the left, center and right superconducting leads.
}
\vspace{-2mm}
\label{fig:sketch}
\end{figure} 

 The $\omega$-SQUIPT is the natural evolution of the standard two-terminal geometry, enriched by a third terminal in the metallic Josephson junction, as sketched in Fig. \ref{fig:sketch}.
It is composed by a T-shaped N nanowire proximized by two S loops, encircling two independent magnetic fluxes.
The  $\omega$-SQUIPT represent a useful tool to explore  the non-trivial physics accessible in multi-terminal Josephson junctions (JJs) in which the Andreev bound states can  cross the Fermi level (zero-energy)~\citep{padurariu_closing_2015} to tailor exotic quantum states~\citep{yokoyama_singularities_2015,riwar_multi-terminal_2015},  or to simulate topological materials able to support Majorana bound states in the case of quasi-ballistic junctions with strong spin orbit coupling \citep{riwar_multi-terminal_2015,van_heck_single_2014}.
 The first $\omega$-SQUIPT has been realized~\citep{strambini} very recently  with a diffusive three-terminal JJ. The experiment, in agreement with theoretical expectations, demonstrates that a superconducting-like gapped state is induced in the weak-link and non-trivially controlled by an external magnetic field. Moreover this state can be topologically classified by the winding numbers of the two S loops.
  
The aim  of this work  is to address  the role of the main experimental parameters of the $\omega$-SQUIPT on the spectral and transport properties. On this purpose, the effects of junction length, transparency of the SN interfaces and inelastic scattering are discussed. In addition to the analysis of the quasi-particle density of states, a study of the supercurrent flowing in the different arms of the device is reported. Such coherent transport properties in the $\omega$-SQUIPT can be a mark of a topological transitions\citep{yokoyama_singularities_2015,riwar_multi-terminal_2015}.  

The manuscript is organized  as follows.
The model based on the solution of the Usadel equation \cite{usadel_generalized_1970, belzig} for  the quasiclassical Green functions formalism is described in Sec. \ref{sec:model}.
The analysis of the local normalized DoS is presented in Sec. \ref{sec:dos} where we discuss the effect of the length of the proximized metallic junction, of the inelastic scattering, and the transparency of the contact interface.
The Josephson and the quasi-particle currents are calculated in Secs. \ref{sec:josephson} and \ref{sec:quasiparticle}, respectively.
In Sec. \ref{sec:conclusion}  we summarize our main findings.

\section{Model and general settings}
\label{sec:model}

The $\omega$-SQUIPT is made of a T-shaped N weak link formed by three diffusive quasi-one dimensional arms of lengths $L_i$ (i=L,C,R), as sketched in Fig. \ref{fig:sketch}. 
Each of the arms is  connected to a superconducting lead $S_i$ with phase  $\varphi_i$ and gap $\Delta_0$.
The three superconducting phases are linked by the two magnetic fluxes $\Phi_L$ and $\Phi_R$ 
piercing the double-loop of the interferometer (see Fig. \ref{fig:sketch}). 
 The properties of the device  can be described by using the isotropic quasi-classical retarded Green function $\hat{g}_i$ which are $2 \times 2$ matrices in the Nambu space \citep{serene_quasiclassical_1983}. In a stationary case  these functions satisfy the Usadel equations in  each arm ($i$) of the $\omega$-SQUIPT ,\citep{usadel_generalized_1970, belzig}
\begin{equation}
\partial_{x} \left(\hat{g}_i \partial_{x} \hat{g}_i \right) +i \frac{(E+i\Gamma _N)}{E_{i}}\left[\hat{\tau}_3 , \hat{g}_i\right]=0 \, \, ,
\label{eq:system1}
\end{equation}
where $\hat{\tau}_3$ is the third Pauli matrix in the Nambu space and $x$ is the normalized spatial coordinate mapping the T-shaped weak link from the center ($x=0$) to the S/N interface ($x=1$). $E_i\equiv \hbar D/L_i ^2$ is the (reduced) Thouless energy associated to each arm of the  link, and $\Gamma _N$ is a parameter that takes into account the inelastic processes in the N region. 
Equation (\ref{eq:system1}) is complemented by the normalization condition
\begin{equation} 
\hat{g}_i^2=\hat{1} \, ,
\label{normalisation}
\end{equation}
and  boundary conditions at the three S/N interfaces and  in the middle of the T-shaped junction.

At the S/N interfaces the Green function has to satisfy the boundary conditions for arbitrary transparency \citep{nazarov_novel_1999,nazarov_blanter_2009}   
\begin{equation}
r_i \,\hat{g}_i \,\partial _x \hat{g}_i =
\frac{2\, [\hat{g}_i,\hat{G}_i]}{4+\tau \left( \left\lbrace \hat{g}_i, \hat{G}_i \right\rbrace -2 \right)} \, ,
\label{eq:nazarov_cond2}
\end{equation}
where $\tau$ is the transmission coefficient, the opacity coefficient $r_i= G_{N_{i}}/G_{B_{i}}$ is the ratio between the conductance of each arm $G_{N_{i}}$ and the barrier conductance $G_{B_{i}}$, and
\begin{equation}
\hat{G}_i = \frac{1}{\sqrt{(E^R)^2-\Delta_0^2}}\left(
\begin{array}{cc}
E^R & \Delta_0e^{i\varphi _i} \\
-\Delta_0 e^{-i\varphi _i} & -E^R 
\end{array}
\right)
\end{equation}
is the BCS Green function of the S$_i$ lead\cite{Bennemann}, $\Delta_0 e^{i\varphi_i}$ is the superconducting order parameter, $E^R\equiv E+i\Gamma _S$, where $\Gamma _S$ is the Dynes parameter \cite{dynes_direct_1978,pekola_dynes}.
Neglecting the inductance of the superconducting loops, we can link the two superconducting phase differences to the two magnetic fluxes: $\varphi_L-\varphi_C= 2 \pi \Phi_L/\Phi_0$ and $\varphi_R-\varphi_C= -2\pi \Phi_R/\Phi_0$, with $\Phi_0=h/2e$ the flux quantum (hereafter $e$ indicates the modulus of the electron charge). 
Notice that for sake of simplicity in Eq.~(\ref{eq:nazarov_cond2}) we have assumed that all the conduction channels at all the interfaces have the same transmission $\tau$ and therefore $G_{B_i}=G_0N_i\tau$, where $G_0$ is the quantum of conductance and $N_i$ the number of conducting channels at the $i$-th interface. 

In the middle of the T-shaped junction, $x=0$, we impose the continuity of $\hat{g}_i$ :
\begin{equation}
\hat{g}_L(x=0) = \hat{g}_C(x=0) = \hat{g}_R(x=0)\, \, ,
\label{eq:continuity}
\end{equation}
and the matrix current conservation
\begin{equation}
\sum _{i=R,C,L} G_{N_i} \, \hat{g}_i \, \partial _x  \hat{g}_i \,\, |_{x=0}\, = 0 \, \, .
\label{eq:inutile}
\end{equation}

In order to solve the Eqs. (\ref{eq:system1}-\ref{eq:inutile}) we introduce  the Riccati parametrization that parametrizes $\hat{g} _i$  in term of two auxiliary functions $\gamma_i(x,E)$ and $\tilde{\gamma}_i(x,E)$.
Therefore equations (\ref{eq:system1},\ref{normalisation}) become a system of six coupled differential equations:
\begin{equation}
\left\lbrace
\begin{aligned}
&\partial ^2 _x \gamma_i-\frac{2\tilde{\gamma}_i}{1+\gamma_i\tilde{\gamma}_i}(\partial_x \gamma_i)^2+2i\left(\frac{E+i\Gamma_N}{E_{i}}\right)\gamma_i=0 \\
&\partial ^2 _x \tilde{\gamma}_i-\frac{2{\gamma}_i}{1+{\gamma}_i\tilde{\gamma}_i}(\partial_x \tilde{\gamma}_i)^2+2i\left(\frac{E+i\Gamma_N}{E_{i}}\right)\tilde{\gamma}_i=0
\end{aligned}
\right. \, \, ,
\label{eq:system2}
\end{equation}
with boundary conditions at  $x=0$ (see  Eqs. (\ref{eq:continuity},\ref{eq:inutile})) (here $i,k \in R, C, L$)
\begin{equation}
\left\lbrace
\begin{aligned}
&\gamma_{i}  = \gamma _{k} \\
&\tilde{\gamma}_{i}  = \tilde{\gamma} _{k} \\
&\sum _i G_{N_i}\,\frac{\partial _x \gamma_i+(\gamma_i)^2\partial _x \tilde{\gamma}_i}{1+\gamma _i \tilde{\gamma}_i}  =0 \\
&\sum _i G_{N_i}\,\frac{\partial _x \tilde{\gamma}_i+(\tilde{\gamma}_i)^2\partial _x {\gamma}_i}{1+\gamma _i \tilde{\gamma} _i}  =0
\label{eq:center_boundary}
\end{aligned}
\right. \, \, .
\end{equation}
At the S/N interfaces ($x=1$) the boundary condition in Eq. (\ref{eq:nazarov_cond2}) reads:
\begin{multline}
r_i \frac{\partial _x \gamma_i + \gamma _i ^2 \partial_x \tilde{\gamma}_i}{(1+\gamma_i \tilde{\gamma}_i)^2} = \\ \frac{(1-\gamma_i \tilde{\gamma}_i)\gamma_i^S - (1-\gamma_i^S \tilde{\gamma}_i^S) \gamma _i}{(1+\gamma _i \tilde{\gamma}_i )(1+\gamma _i^S \tilde{\gamma}_i^S) -\tau (\gamma_i^S - \gamma_i)(\tilde{\gamma}_i^S - \tilde{\gamma}_i)}\; ,\label{BC_finiteR}
\end{multline}
and an analogous equation after substituting $\gamma_i$ by $\tilde{\gamma}_i$. 
The functions ${\gamma}^S_i = \gamma_0 e^{-i \phi_i}$, $\tilde{\gamma}^S _i = -\gamma_0 e^{i \phi_i}$ are the auxiliary functions parametrising the BCS bulk Green functions, with
\begin{equation}
\gamma_0 = \frac{-\Delta_0}{E+i\Gamma_S+i\sqrt{(\Delta_0)^2-(E+i\Gamma_S)^2}}\, .
\end{equation}

By solving these equations numerically, we obtain the functions $\gamma_i$, that determine 
 the DoS , the supercurrent and the quasiparticle current  in the $\omega$-SQUIPT.  All these observables are discussed in the next sections.

In the following calculations, we assume a full symmetric structure, i.e. $L_L=L_C=L_R\equiv L$ and $G_{N_L}=G_{N_C}=G_{N_R}\equiv G_N$; thus, we define a single Thouless energy for the whole junction: $E_{Th}\equiv\hbar D/ (2L)^2=E_i/4$, to adopt the same energy scale defined in two-terminal geometry. When not explicitly indicated we will assume ideal interfaces and hence impose the continuity of $\gamma$ at the S/N interfaces. Only when analyzing the role of the S/N interfaces resistances we will make use of boundary condition (\ref{BC_finiteR}).

\section{The density of states in the N region} 
\label{sec:dos}
In this section we investigate the DoS in the T-shaped normal region and its dependence on various parameters. 
The local normalized DoS in the $i$-th arm of the proximized nanowire is given by:
\begin{multline}
N_i (x,E,\Phi_L,\Phi_R)=
\frac{1}{2}\operatorname{Re} \operatorname{Tr} \left\lbrace \hat{\tau} _3 \hat{g}_i \right\rbrace= \\
= \operatorname{Re} \left\lbrace \frac{1-\gamma_i\tilde{\gamma}_i}{1+\gamma_i \tilde{\gamma}_i} \right\rbrace \, \, .
\label{eq:DoS_formula}
\end{multline}
\begin{figure}[t!]
\includegraphics[width=8.5cm]{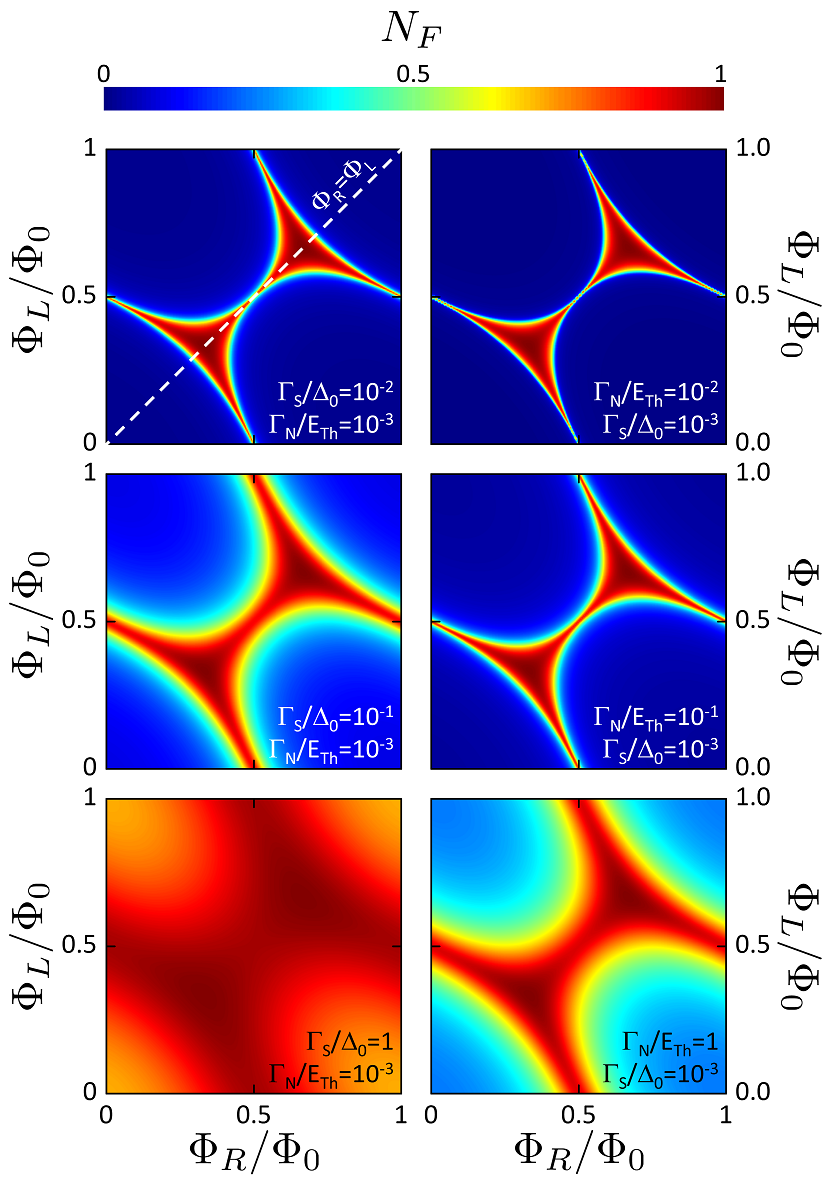}
\vspace{-0mm} 
\caption{ (Color online) Evolution of the DoS at Fermi energy $N_F(\Phi_L,\Phi_R)$, for increasing pair-breaking scattering both in the S leads $\Gamma _S$ (left column) and in the N weak link $\Gamma _N$ (right column). The values of $\Gamma_N/E_{Th}$ and $\Gamma_S/\Delta_0$ are reported in each panel. The weak link is of an intermediate length ${E}_{Th}/\Delta_0=0.5$ and the S/N interfaces are transparent.}
\label{fig:dyn_VS_in}
\vspace{-2mm}
\end{figure}
We start by analyzing the local DoS at the Fermi level in the middle of the T-shaped N wire, $N_F (\Phi_L, \Phi_R) \equiv N_i(x=0,E=0,\Phi_L,\Phi_R)$ as a function of 
the two fluxes $\Phi_{L}$ and $\Phi_R$  through the two loops. 
Fig. \ref{fig:dyn_VS_in} shows a typical result for this dependence.
 We clearly identify gapped (in blue) regions separated by gapless ones (in red). From the top panel to the bottom one, it is noticeable the effects of finite quasi-particle lifetime in the superconductor leads  (left column) and inelastic scattering in the normal metal (right column), described respectively by the parameters $\Gamma_S/\Delta_0$ and $\Gamma_N/E_{Th}$.

It is instructive to note that  the density of states precisely at Fermi energy does not depend on 
the size of the normal region, unless we assume a significant rate of inelastic scattering $\Gamma_N$. In the latter case, the size enters the equations through the ratio $\Gamma_N/E_{Th}$.

The white dashed line tracks the case of equal fluxes in the two loops, $\Phi_L=\Phi_R\equiv\Phi$, experimentally realizable placing a symmetric $\omega$-SQUIPT in a homogeneous magnetic field. Figure \ref{fig:dyn_VS_in} suggests that the gap closes at $\Phi\approx\Phi_0/3$, as confirmed by recent measurements\citep{strambini}.
Interestingly enough, to each gapped region, it can be assigned a topological index defined by the pair of numbers obtained by the integration of superconducting phase gradient over the left and right loop\cite{strambini}. 
We note that, our results well agree with the recent findings of Ref.\cite{linder_2016}, where an analytical approach for a multi-terminal geometry at the Fermi level has been investigated. 

\begin{figure}[t!]
\includegraphics[width=8.5cm]{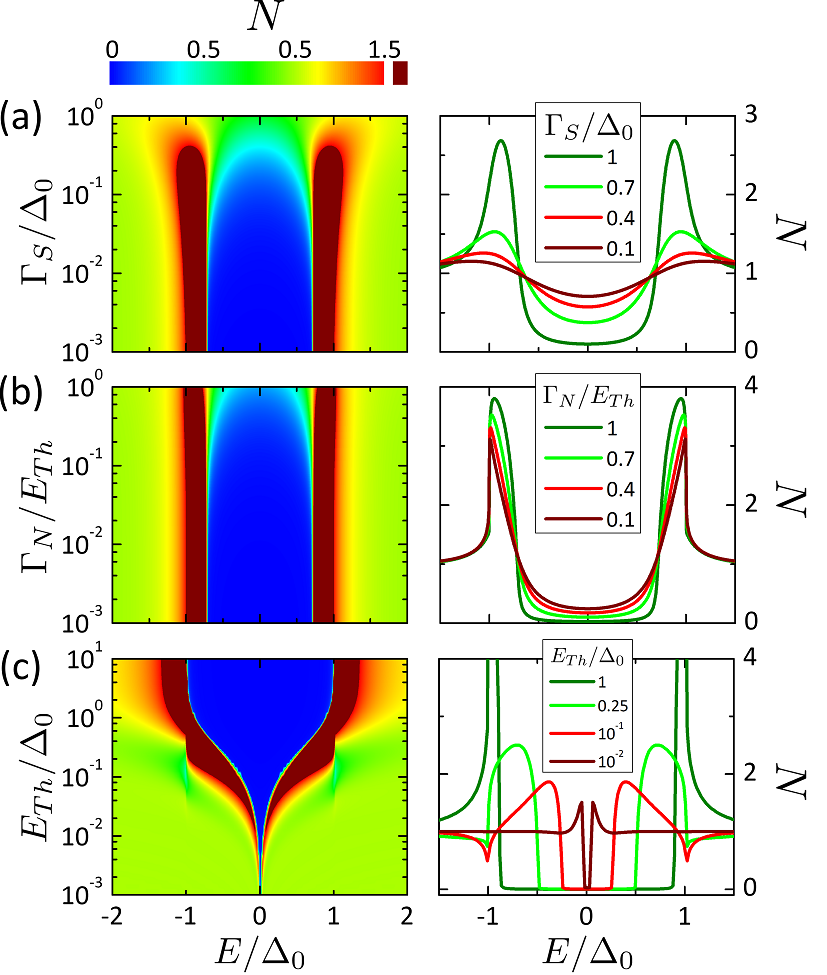}
\vspace{-0mm} 
\caption{(Color online) DoS in the center of the three-terminal junction ($x=0$) calculated at zero fluxes, $\Phi_L=\Phi_R=0$.
(a) Dependence of the DoS on $\Gamma _S/\Delta_0$(fixed $E_{Th}/\Delta_0 = 0.5$ and $\Gamma _N/E_{Th}=10^{-3}$). 
(b) Dependence of the DoS on $\Gamma _N/E_{Th}$ (fixed $E_{Th}/\Delta_0 = 0.5$ and $\Gamma _S/\Delta_0=10^{-3}$).
(c) Dependence of the DoS on the Thouless energy $E_{Th}/\Delta_0$ (fixed $\Gamma _S =\Gamma _N = 10^{-3}\Delta_0$).
\label{fig:DoSeS}} 
\vspace{-2mm}
\end{figure}

We consider now the DoS at equal fluxes for all energies. In Fig. 3 we compare the detrimental role played by $\Gamma_S$, $Gamma_N$ and $E_{Th}$ in the DoS calculated at $\Phi=0 $ for which the proximity effect is maximized. The main common feature is the appearance of an induced minigap $\Delta _w$.
As expected, increasing $\Gamma_N$ or $\Gamma _S$ causes the smearing of the gapped feature, as one can see in panel (a) and (b) of Fig. \ref{fig:DoSeS}. The dependence on Thouless energy (then on junction size) is showed in the panel (c) of Fig. \ref{fig:DoSeS}. Similarly to two-terminal geometry the induced minigap $\Delta_w$ decreases with decreasing Thouless energy \cite{hammer_density_2007}.

\begin{figure}[t!]
\includegraphics[width=8.5cm]{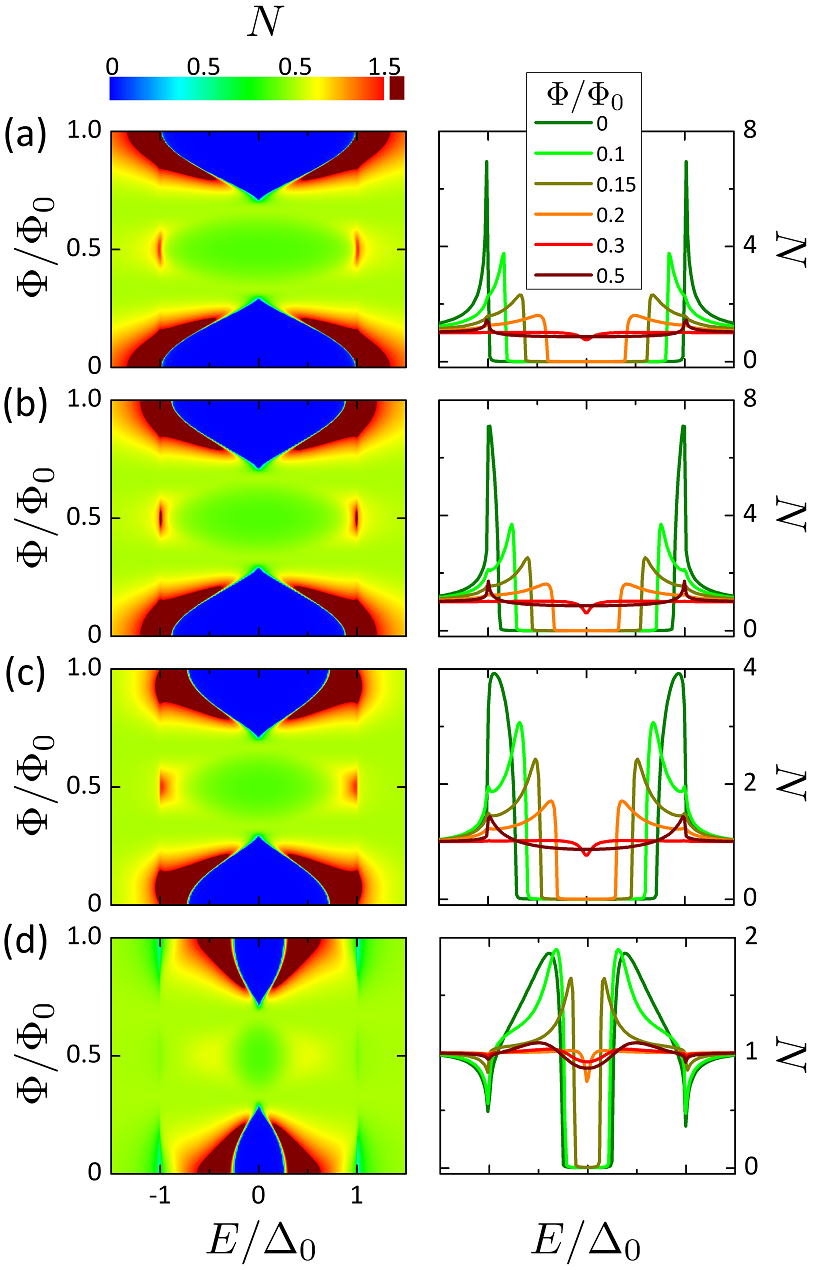}
\vspace{-0mm} \caption{(Color online) DoS calculated in the middle of the three-terminal junction ($x=0$) for equal fluxes $\Phi_{L}=\Phi_{R}\equiv\Phi$ with $\Gamma_S=\Gamma_N=10^{-3}\Delta_0$. 
Each panel corresponds to a different Thouless energy:
(a) $E_{Th}/\Delta_0=5$;
(b) $E_{Th}/\Delta_0=1$; 
(c) $E_{Th}/\Delta_0=0.5$; 
(d) $E_{Th}/\Delta_0=0.1$.
\label{fig_DoSes}} \vspace{-2mm}
\end{figure}

In Figure \ref{fig_DoSes} we illustrate the dependence of the DoS on equal magnetic fluxes $\Phi=\Phi_R=\Phi_L$. Each panel corresponds to a different length. From top to bottom we explore the behavior of the DoS from short to long junctions, with $E_{Th}/\Delta_0 = 5, 1, 0.5, 0.1$, respectively. 
In the  short-junction limit (Fig. \ref{fig_DoSes}(a)) our results are in good agreement with those of Ref. \onlinecite{padurariu_closing_2015}, obtained within the circuit theory. This limit is achievable for conventional metals in use in nanofabrication at $L\lesssim 100$ nm.
Above this limit the minigap rescales in energy (as observed also in Fig. 3 (c)) while the behavior in $\Phi$ is practically unaffected. In fact for all the lengths explored the induced minigap is modulated by the magnetic flux and disappear in an extended flux interval  $1/3<\Phi/\Phi_0<2/3$, repeated with $\Phi_0$ periodicity. This continuous gapless region is the main hallmark of multi-terminal JJs (recently observed experimentally in Ref. \onlinecite{strambini}) and it is a consequence of the crossing of the Andreev bound states at zero energy.

\begin{figure}[t!]
\includegraphics[width=8.5cm]{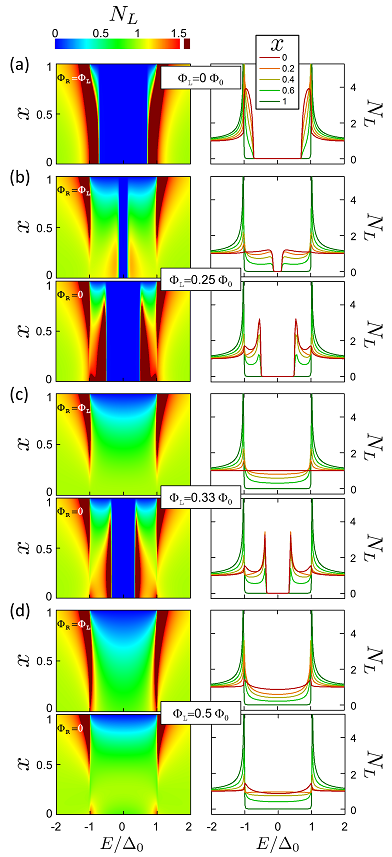}
\vspace{-0mm} \caption{(Color online)
Spatial dependence of the DoS evaluated at different fluxes, with $E_{Th}/\Delta_0=0.5$ and $\Gamma_S=\Gamma_N=10^{-3} \Delta_0 $. Each central box indicates the value of the flux $\Phi_{L}$ associated to the near plots; the top plots show the case of equal fluxes $\Phi_{R}=\Phi_{L}$ and the bottom plots show the single flux case with $\Phi_{R}=0$. (a) $\Phi_L=0$; (b) $\Phi_L=0.25\Phi_0$; (c) $\Phi_L=0.33\Phi_0$; (d) $\Phi_L=0.5\Phi_0$. 
} \vspace{-2mm}
\label{fig:spatial_dependence_one_flux}
\end{figure}

We now discuss the spatial dependence of the DoS along the N region. This point is very relevant for two main reasons. 
From a practical point of view, in order to simulate realistically the differential conductance of a tunnel contact between the weak link and the probe, the DoS needs to be averaged over the  contact area (see Sec. \ref{sec:quasiparticle} below). From a more fundamental aspect, it is important to understand whether the gapped regions in  Fig. \ref{fig:dyn_VS_in} are a nonlocal property of the junction, as already proved experimentally for the minigap in two-terminal SNS junctions \cite{le_sueur_phase_2008}. 

Figure \ref{fig:spatial_dependence_one_flux} shows the dependence of the DoS on $x$ in the left arm, i.e. $N_L $. 
Due to the continuity imposed at the S/N interfaces ($x=1$), the DoS is here equal to its BCS value and there is no modulation with the magnetic flux. 
Inside the N region the DoS evolves with a well defined minigap $\Delta_w$ which  is constant in the whole T-shape region.  Whereas the minigap is a non-local property that can be modulated  by the magnetic fluxes , the shape of the DoS for energies larger than the minigap changes along the junction. Notice that  for a single flux ($\Phi_R=0$) the DoS shows two additional peaks at the minigap of the nanowire at energy $\pm\Delta_w$ similar to the edge peaks expected in two-terminal SNS junctions\citep{zhou_density_1998}. 

\begin{figure}
\includegraphics[width=8.5cm]{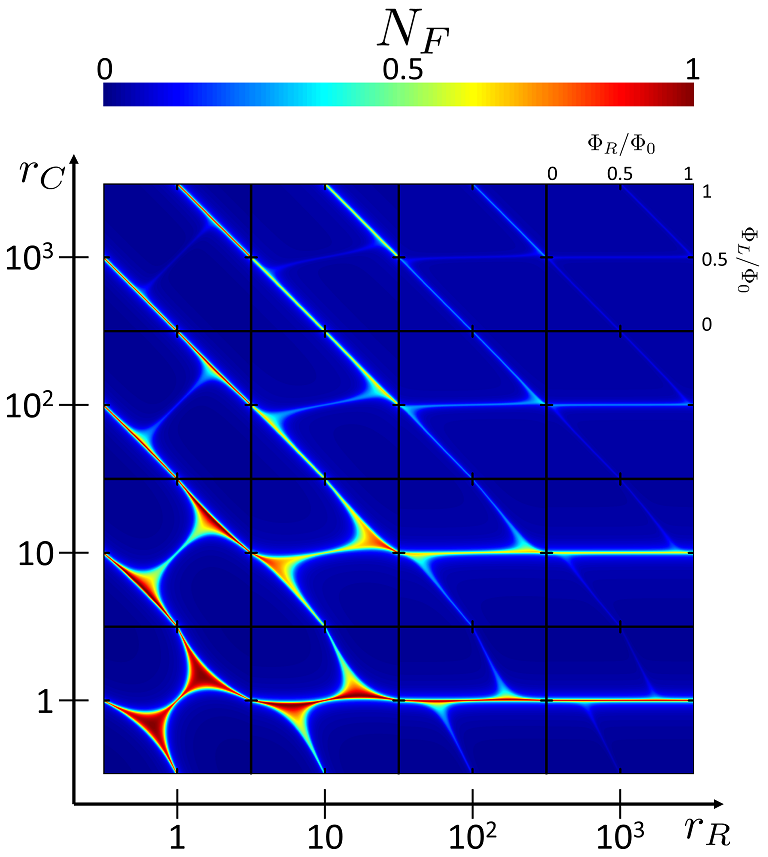}
\vspace{-0mm} \caption{(Color online) 
Evolution of the DoS at the Fermi energy $N_F(\Phi_L,\Phi_R)$ for different values of S/N interface opacities $r_R$, and $r_C$ reported in the x and y axis respectively. Here $r_L=1$, $E_{Th}/\Delta_0=0.5$ and  $\Gamma_S=\Gamma_N= 10^{-3} \Delta_0$.
}
\label{fig:entomology} \vspace{-2mm}
\end{figure} 

We finally  concentrate on the role of the S/N interface resistances in the energy spectrum of the DoS. 
These resistances are encoded in the three opacity parameters $r_i$ defined in Eq~\ref{eq:nazarov_cond2}. 
The increasing of the opacity of all the interfaces weakens the proximity effect in the JJ, which in turns is reflected in an effective reduction of the minigap\citep{hammer_density_2007}.
In Figure \ref{fig:entomology} we show  $N_F(\Phi_{L},\Phi_{R})$   for different values of $r_C$ and $r_R$, and by  keeping $r_L=1$. 
In the symmetric case, $r_R=r_C=1$, we obtain the symmetric "butterfly" shape observed in Fig. \ref{fig:dyn_VS_in} for ideal interfaces. 
Asymmetries in the interface transparencies leads to an  asymmetric configuration of the gapped states  in the two-flux space.
This asymmetry can be understood by considering three limiting cases:
(i) When the right terminal is almost disconnected to the system, $r_R \gg (r_C,r_L)$ (bottom-right plot), $\Phi_R$ does not drive the state of the JJ. The latter effectively behaves as a two terminal junction in which the gapless state is punctual in the flux $\Phi_L$ that controls the proximity effect in the junction. 
(ii) Similarly when $r_C \gg (r_R,r_L)$ (top left plot), the central terminal is disconnected and the proximity effect in this two-terminal JJ is controlled by the total flux in the interferometer $\Phi_L+\Phi_R$.
(iii) When both the interfaces are opaque $r_R=r_C \gg r_L$ (top right panel), both $\Phi_L$ and $\Phi_R$ do not drive the proximity effect. In the weak link, a non-modulated gapped state is induced by the contact with the left S/N interface.

\begin{figure}[t!]
\includegraphics[width=8.5cm]{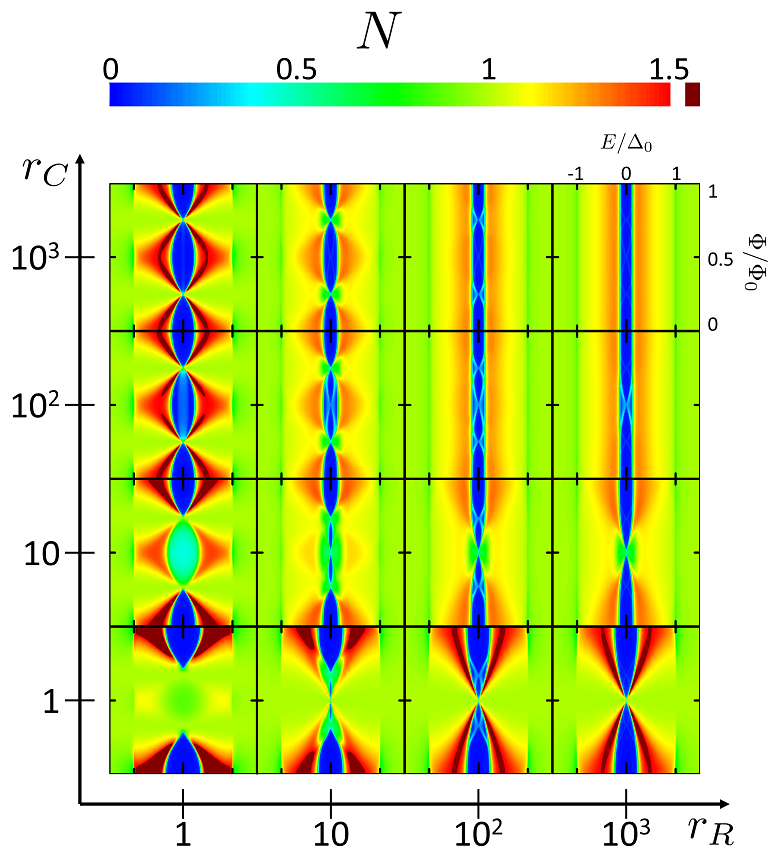}
\vspace{-0mm} \caption{(Color online) Energy spectrum of the DoS  as a function of equal fluxes $\Phi_{L}=\Phi_{R}\equiv \Phi $ and calculated for different values of S/N interface opacity $r_R$, and $r_C$ reported in the x and y axis, respectively. Here $r_L=1$, $E_{Th}/\Delta_0=0.5$ and  $\Gamma_N=\Gamma_S= 10^{-3} \Delta_0$.
\label{fig1:interfaces}} \vspace{-2mm}
\end{figure}

Finally in Figure \ref{fig1:interfaces}  we show the full  energy spectrum of the DoS, in the equal fluxes configuration $\Phi_R=\Phi_L\equiv\Phi$. 
 It is worth noting that small asymmetries in the interface resistances can generate a second small gapped region at $\Phi=\Phi_0/2$, an additional feature that have been observed experimentally\citep{strambini}.

\section{Josephson current}
\label{sec:josephson}
The presence of finite magnetic fluxes $\Phi_L$ and $\Phi_R$, leads to  supercurrents flowing  in the proximized metallic nanowire. These supercurrents have a variety of physical behaviors depending on the junction characteristics \cite{golubov2004, likharev}.
Within the quasiclassical theory, the supercurrent flowing in  the $i$-th arm of the $\omega$-SQUIPT can be written as
\begin{equation}
\mathcal{I}_{i} = \int _{-\infty} ^{+\infty} \tanh \left( \frac{E}{2k_B T}\right) S_i(E)  dE \, \, ,
\end{equation}
where $T$ is the temperature, $k_B$ is the Boltzmann constant and $S_i(E)$ is the outgoing spectral supercurrent density in the $i$ arm
\begin{multline}
S_i(E)=-\frac{G_{N_i}}{4e}\operatorname{Re}\left\lbrace\operatorname{Tr}\left\lbrace\hat{\tau}_3\hat{g}_i\partial_x\hat{g}_i\right\rbrace\right\rbrace=\\=\frac{G_{N_i}}{e} \operatorname{Re} \left\lbrace \frac{\tilde{\gamma}_i \partial _x \gamma _i - \gamma _i \partial _x \tilde{\gamma}_i}{(1+\gamma _i \tilde{\gamma}_i)^2 } \right\rbrace \, \, .
\end{multline}

In this section we investigate the outgoing supercurrent flowing through the different arms of the device and its dependence on the magnetic fluxes $\Phi_R$ and $\Phi_L$, for transparent S/N interfaces. At first, we consider the simple case of equal magnetic fluxes $\Phi_R=\Phi_L\equiv \Phi$. In this case, for symmetry reasons, there is no supercurrent flowing through the  central arm ${\cal I}_C=0$, and thus due to current conservation one has ${\cal I}_L = - {\cal I}_R$. Physically this means that there is a supercurrent that flows from the right arm to the left one.
\begin{figure}[t!]
\includegraphics[width=8.5cm]{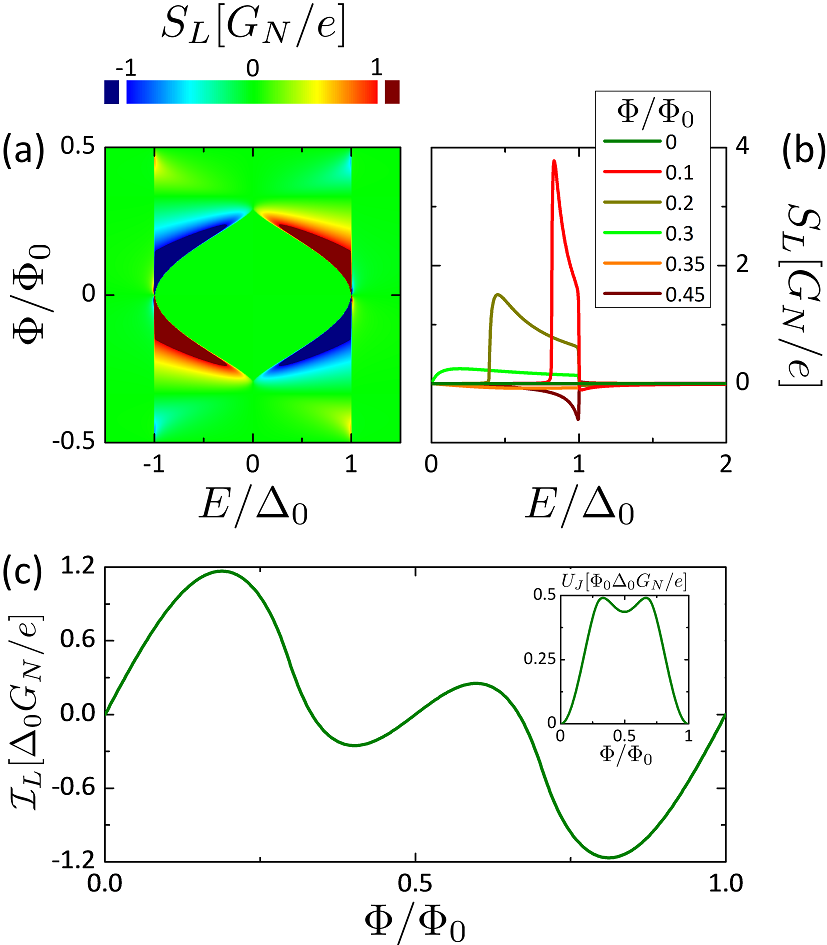}
\vspace{-0mm} 
\caption{(Color online) Outgoing supercurrent of the left arm in the case of equal fluxes $\Phi_R=\Phi_L=\Phi$.
Panel (a) supercurrent spectral density $S_L(E)$ in the case of $E_{Th}/\Delta_0=5$ and with $\Gamma_S=\Gamma_N=10^{-3}\Delta_0$. Related cuts at different fluxes $\Phi/\Phi_0$ are reported in panel (b). Panel (c) shows the supercurrent $\mathcal{I}_L$ at a fixed temperature $T=0.02 \, T_C$. $\mathcal{I}_L$ has a periodic behavior as a function of $\Phi$, with nodes due to the three-terminal junction at $\Phi/\Phi_0=0,\,1/3,\,1/2,\,2/3$; see also the cuts present in panel (b).
}
 \label{fig:supercurrent_l}
\vspace{-2mm}
\end{figure}
We analyze this quantity in Fig. \ref{fig:supercurrent_l}, showing the supercurrent ${\cal I}_L$ and its spectral density $S_L(E)$ for the left arm, at a fixed temperature $T=0.02 \, T_c$.
The supercurrent spectral density $S_L(E)$, present in Fig. \ref{fig:supercurrent_l} (a), strongly resembles the quasi-particle DoS, specifying the distribution on energy of Andreev-bound states which carry the supercurrent. In Fig. \ref{fig:supercurrent_l}(a), where we plot a representative example with $E_{Th}/\Delta_0=5$, one can see that most of the distribution takes place below the superconducting gap $\Delta_0$. Above it there is an evanescent contribution that brings a counterflowing current, which results in a reduction of the critical current. We note that, for shorter junctions, which corresponds to larger value of $E_{Th}/\Delta_0$, the number of states below the superconducting gap increases, giving a greater contribution to supercurrent.

Looking at the color plot in Fig. \ref{fig:supercurrent_l} (a) and the energy cuts in Fig. \ref{fig:supercurrent_l} (b), a change of sign at all energies is evident for $\Phi/\Phi_0=1/3$. This particular value of the flux correspond exactly to the one in which there is a transition from a gapped to a gapless state in the DoS, see Fig. \ref{fig_DoSes}. As for the DoS, this feature does not depend on the junction length. Importantly, this suggests that the supercurrent can be  an alternative hallmark of a topological transition in the three-terminal JJ. This characteristic at $\Phi/\Phi_0=1/3$ is indeed reflected in the supercurrent $\mathcal{I}_S$ as shown in Fig. \ref{fig:supercurrent_l}(c).\\
 To better understand the behavior of ${\cal I}_S$, we can consider the simple case in which the Usadel equations (\ref{eq:system2}) can be linearized. Although this is fully justified in the case of weak proximity effect, with very opaque S/N interfaces ($\tau\ll 1$ and $R_i\gg 1$), it allows for an analytic solution of the system of differential equations (\ref{eq:system2}). As we now discuss, this approach can reproduce most of the qualitative features present in Fig. \ref{fig:supercurrent_l}(c). In particular, for equal interfaces and arm lengths one obtains
\begin{equation}
\mathcal{I}_L = I_0 \left[
\sin \left( 2\pi\frac{2\Phi}{\Phi_0} \right) + \sin \left( 2\pi\frac{\Phi}{\Phi_0} \right)
\right]
\label{eq:analytical_curr}
\end{equation}
where $I_0$ represents the critical current, whose precise value can be calculated using the linearized Usadel equation \cite{hammer_density_2007}.
Equation (\ref{eq:analytical_curr}) is a periodic function of $\Phi$ with period $\Phi_0$ and it presents nodes at $\Phi/\Phi_0=0,\, 1/3,\, 1/2,\, 2/3$. This corresponds to the behavior of the supercurrent shown in Fig. \ref{fig:supercurrent_l}, where ${\cal I}_L$ is evaluated with a full numerical solution of the Usadel equation (without any weak-proximity assumption).
The fact that a linear approach well reproduces most of the qualitative features present in the general case is tightly connected to the three-terminal geometry and to its topological properties. In particular, it indicates that these phase-features on the Josephson currents are robust against imperfections and possible microscopic details.
It is interesting to notice that, even if in the equal fluxes case there is a supercurrent flow in the side arms and no supercurrent in the central arm, the behavior is not analogous to a two-terminal JJ linked to a loop with a total flux $2\Phi$. This can be inferred from the additional node present at $\Phi/\Phi_0=1/3$. To underline this we can consider the Josephson energy of the junction $U_J$. In full analogy with the two-terminal expression, it reads
\begin{equation}
U_J = \frac{\Phi_0}{2\pi}\int\mathcal{I}_L d(\phi_L-\phi_R)= 2\int \mathcal{I}_L d\Phi \, \, .
\label{eq:U_J}
\end{equation}
This quantity is reported in the inset of Fig. \ref{fig:supercurrent_l}(c), for $E_{Th}/\Delta_0=5$.
The Josephson energy $U_J$ has two minima at $\Phi/\Phi_0=0$ and $\Phi/\Phi_0=1/2$; the global minimum is $\Phi/\Phi_0=0$ as in the two-terminal case. The presence of additional local minima is a peculiar feature of the three-terminal JJ and is not present in a two-terminal one. 
A junction with such a behavior is called sometimes in the literature a 0' junction\cite{zero_prime_junc,Rozhkov2}, due to the presence of metastable states related to the local minima at $\Phi/\Phi_0=1/2$. 
The maximum at $\Phi/\Phi_0=1/3$ determines the node present in the supercurrent. The presence of this local minima is a direct consequence of the non-trivial topological configuration which can be achieved with the $\omega$-SQUIPT and is associated to the presence of the central arm in this three-terminal configuration.
\begin{figure}[t!]
\includegraphics[width=8.5cm]{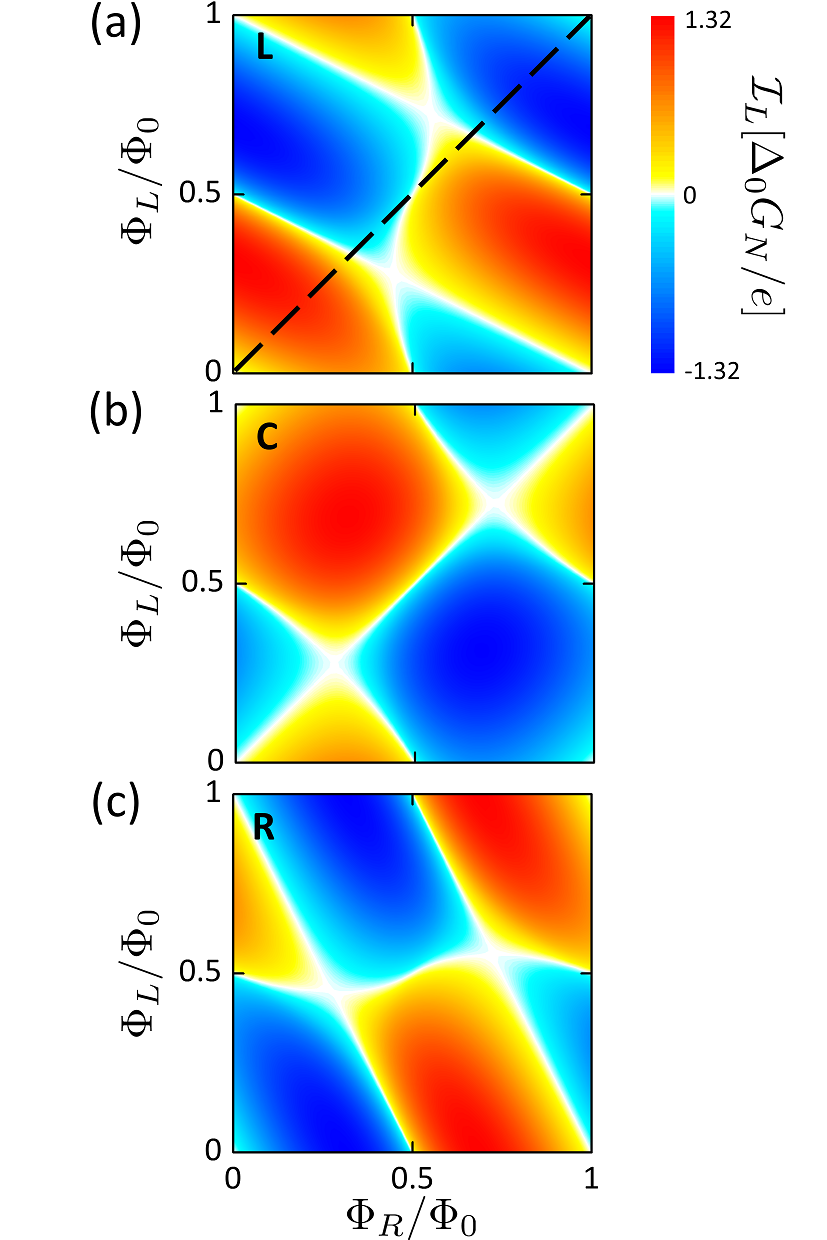}
\vspace{-0mm} 
\caption{(Color online)
Color plot of the supercurrents in each arm for different magnetic fluxes $\Phi_R$ and $\Phi_L$. Here we have fixed $E_{Th}/\Delta_0=5$ and $\gamma_S= \gamma_N=10^{-3}\Delta_0$, and $T=0.02 T_c$. The supercurrent flowing out of the left, central and right arm are plotted in the panel (a), (b), and (c), respectively.
}
\label{fig:supc_map}
\vspace{-2mm}
\end{figure}

Let us now discuss the case of different fluxes $\Phi_L$ and $\Phi_R$ and their influence on the $i$-th arm supercurrent. The outgoing supercurrents ${\cal I}_L$, ${\cal I}_C$, and ${\cal I}_R$ are reported in the three panels of Fig. \ref{fig:supc_map} for transparent S/N interfaces and for  fixed parameters $E_{Th}/\Delta_0=5$, $\Gamma_N=\Gamma_S=10^{-3}\Delta_0$, and temperature $T=0.02\,T_c$. The dashed line in panel (a) correspond to the panel (c) in Fig. \ref{fig:supercurrent_l}. We immediately note that, in the general case with $\Phi_L\neq\Phi_R$ a finite supercurrent is flowing out of the central arm. As one would expect, the three quantities are not independent, but they are related by current conservation, i.e. $\sum_{i=L,C,R} {\cal I}_i=0$. As before, the qualitative behavior and the main features present in Fig. \ref{fig:supc_map} can be understood inspecting the solution of the linearized Usadel equations.
In this case, the supercurrent in each arm is the superposition of the circulating supercurrent in each loop, that gives
\begin{equation}
\label{eq:spc}
\begin{aligned}
&\mathcal{I}_L= I_0 \left[
\sin\left(2\pi\frac{\Phi_L+\Phi_R}{\Phi_0}\right)+ \sin\left(2\pi\frac{\Phi_L}{\Phi_0}\right)
\right]\\
&\mathcal{I}_C= I_0 \left[
\sin\left(2\pi\frac{\Phi_R}{\Phi_0}\right)- \sin\left(2\pi\frac{\Phi_L}{\Phi_0}\right)
\right]\\
&\mathcal{I}_R= -I_0 \left[
\sin\left(2\pi\frac{\Phi_L+\Phi_R}{\Phi_0}\right)+ \sin\left(2\pi\frac{\Phi_R}{\Phi_0}\right)
\right]
\end{aligned}
\end{equation}
Again, these simple analytical expressions well reproduce the periodic behavior and the shape of the supercurrents shown in Fig. \ref{fig:supc_map}. The full numerical solution extends beyond the linear approximation, which is not able to capture the correct value of the critical current and other details. However, the periodicity and the presence of nodes at precise values of $\Phi_{L,R}/\Phi_0$ are well reproduced by Eq. (\ref{eq:spc}). This fact corroborate the idea that these features are robust in a topological sense and connected to the non-trivial geometry of the three-terminal JJ.

\section{Magnetometric characteristics of the $\omega$-SQUIPT}
\label{sec:quasiparticle}

As shown in Sec. \ref{sec:dos}, the DoS in the junction is modulated by the magnetic fluxes piercing the superconducting loops of the $\omega$-SQUIPT. The transport properties of quasi-particle in the junction can be tuned from metallic-like (in gapless state) to insulating-like (in gapped state). As a consequence, the electrical conduction through the tunnel barrier between the junction and the probe (Fig. \ref{fig:sketch}) is altered\cite{giaever1960,tinkham1972}. This effect allows to perform magnetometric measurement. In a two-terminal SQUIPTs an high sensitivities have been demonstrated\cite{giazotto_superconducting_2010,ronzani_highly_2014}. In the following we evaluate the sensitivity of the $\omega$-SQUIPT.

Considering a tunnel probe placed in the middle of the T-shaped N region and covering each arm by a length $l_i$, the electrical characteristics depend on the spatial average of the local DoS $N_i(x, E,\Phi_L,\Phi_R)$ over the contact area, given by\cite{tinkham1972}:
\begin{equation}
\bar{N}(E, \Phi_L,\Phi_R)\equiv \sum\limits _{i=R,C,L} \frac{1}{w_i}\int ^{w_i} _0 N_i(x,E ,\Phi_L, \Phi_R) dx\, \, ,
\end{equation} 
where $w_i=l_i/L_i$. By applying a voltage $V$ between the tunnel probe and the junction, a finite tunneling current flows through the contact, whose expression reads
 \begin{multline}
I=\frac{1}{e R_t} \int N_P(E-eV)\bar{N}(E) \times \\ \times [f_F(E)-f_F(E-eV)] dE \, \,  , 
\end{multline}
where $R_t$ is the resistance of the tunnel contact, $f_F(E)$ indicates the equilibrium Fermi-Dirac distribution and $N_P(E)$ is the probe DoS.
Like in the usual SQUIPT~\citep{giazotto_superconducting_2010,dambrosio_normal_2015}, the metallic probe can be made 
of a normal or superconducting material. These two cases are denoted  in the following as normal probe (NP) 
or superconducting probe (SCP), whose normalized DoS are respectively given by $N_P(E)=1$ and
\begin{equation}
N_P(E)= \left| \operatorname{Re} \frac{E + i\Gamma_2 }{\sqrt{(E + i\Gamma_2)^2 - \Delta_2(T)^2}} \right|\; .
\end{equation}
Here $\Gamma_2$ and $\Delta_2$ indicate the Dynes parameter and  gap of the superconducting probe.
 In general,  $\Gamma_2$ and $\Delta_2$ parameters can be different to those of the  $\omega$-SQUIPT described so far.  For sake of simplicity, we assume  that $\Gamma_2=\Gamma_S=10^{-4}\Delta_0$ and $\Delta_2=\Delta_0$ and choose $T=0.02\, T_c$.  We consider two symmetric $\omega$-SQUIPT :  one with $E_{Th}=0.5 \Delta_0$ and $w_i=0.2$ and a second one with $E_{Th}=5 \Delta_0$ and $w_i=0.68$, for $i=L,R,C$.  For a  Al-Cu based device, these  parameters correspond to $L_i\approx 90\,\si{\nano\meter}$ and $L_i\approx 30\, \si{\nano\meter}$ respectively and a contact length in each arm $l_i\approx 20 \, \si{\nano\meter}$. All these values are achievable with state-of-the-art nanofabrication techniques\cite{rabani2009,ronzani_highly_2014}.

\begin{figure}[t!]
\includegraphics[width=8.5cm]{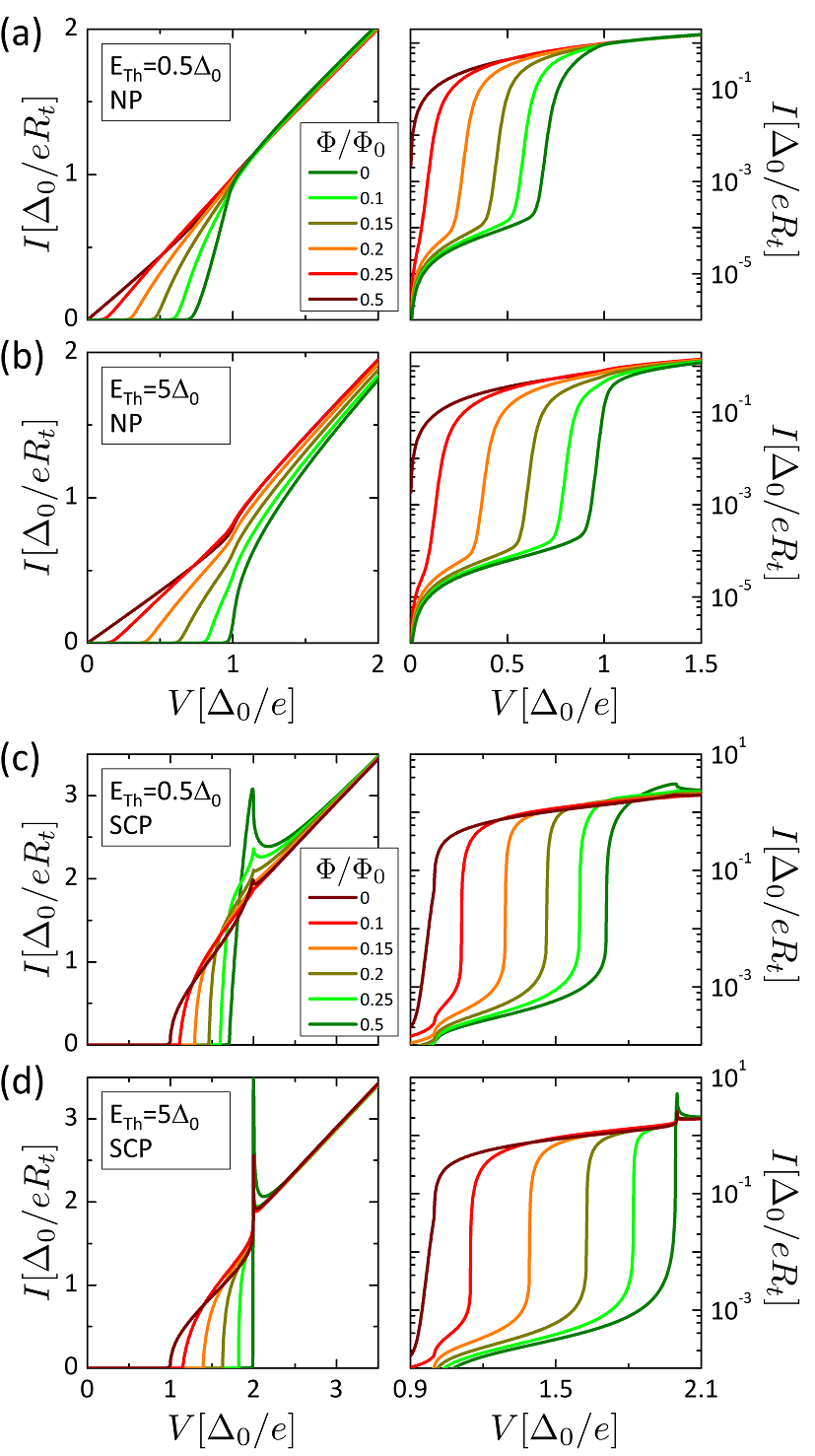}
\vspace{-0mm} \caption{
(Color online) I-V characteristics of the tunnel contact between the probe and the $\omega$-SQUIPT junction at different values of flux $\Phi$, with ideal S/N interfaces.
Here, $\Gamma_S=\Gamma_N= 10^{-4}\Delta_0 $. This quantity is reported both in linear (left column) and Log $y$ (right column) scale.
panel (a), (b) refer to the $\omega$-SQUIPT with a normal metallic probe NP and respectively $E_{Th}/\Delta_0=0.5$ and $E_{Th}/\Delta_0=5$. panel (c),(d) refer to the $\omega$-SQUIPT with a superconducting probe SCP and $E_{Th}/\Delta_0=0.5$ and $E_{Th}/\Delta_0=5$, respectively. 
\label{fig:IV}} \vspace{-2mm}
\end{figure}
\begin{figure}[t!]
\includegraphics[width=8.5cm]{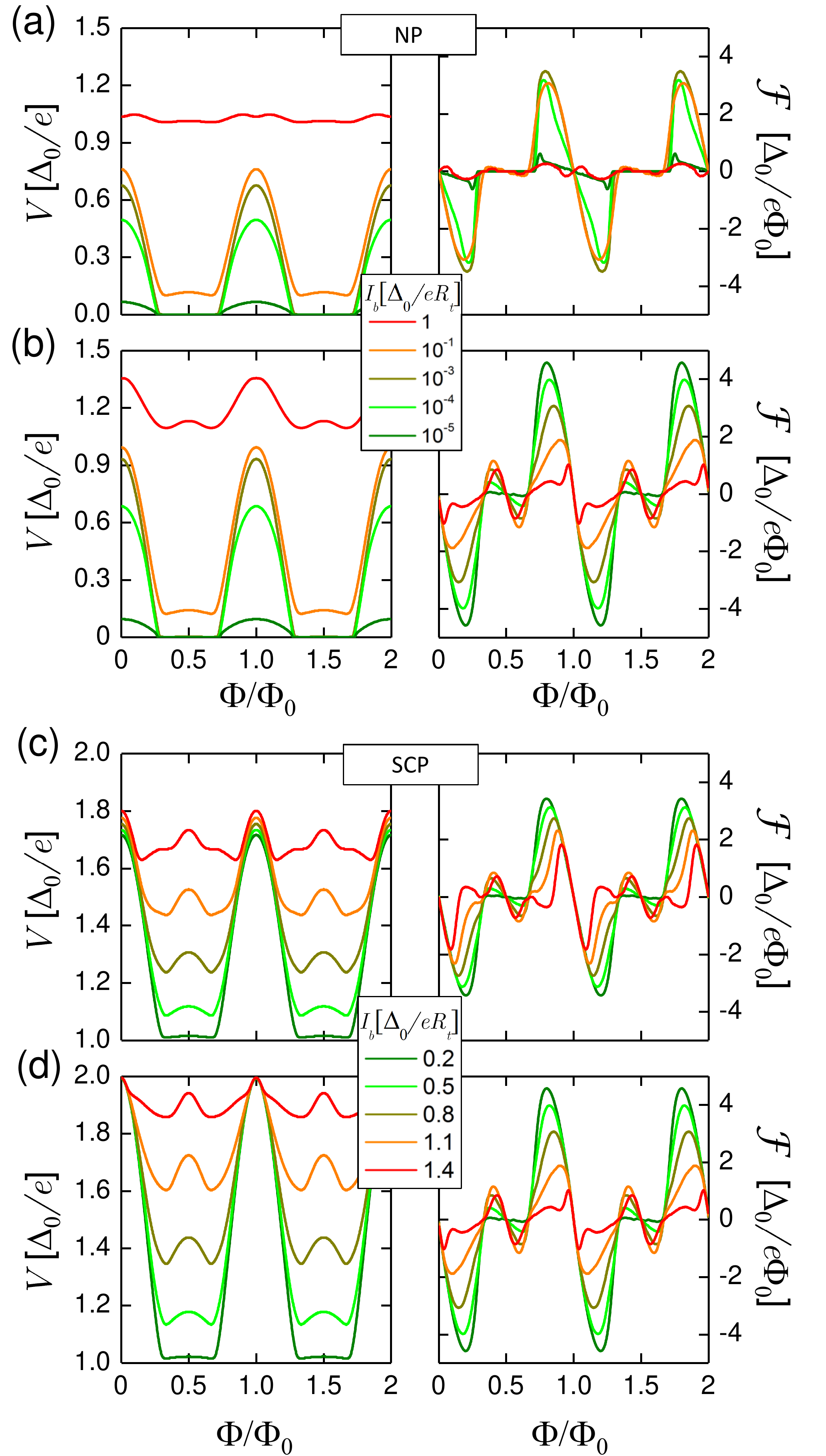}
\vspace{-0mm} \caption{(Color online) { Left column:} Flux to voltage characteristic $V_I(\Phi)$ of the $\omega$-SQUIPT. { Right column:} Transfer function $\mathcal{F}$ associated to the flux Voltage characteristics. Here, the S/N interfaces are transparent and $\Gamma_S=\Gamma_N= 10^{-4}\Delta_0$. (a), (b)  Plots correspond to the case of $\omega$-SQUIPT with a normal probe NP and respectively $E_{Th}/\Delta_0=0.5$ and $E_{Th}/\Delta_0=5$. (c), (d) Plots refer to the $\omega$-SQUIPT with a superconducting probe SCP and respectively $E_{Th}/\Delta_0=0.5$ and $E_{Th}/\Delta_0=5$.
\label{fig:F2V_SCP}} \vspace{-2mm}
\end{figure}

Figure \ref{fig:IV} shows the current-voltage (I-V) characteristic in linear and logarithmic scale for fluxes $\Phi$. panel (a) and (b) refer to the NP case, while (c) and (d) refer to SCP.
The main modulation in the I-V characteristic happens in the flux interval $\Phi/\Phi_0=[0, 1/3]$, for which the weak link is in the gapped state. \
The main differences between the NP and SCP is the presence in the latter of a permanent voltage gap and peaks due to the superconducting probe. The Y-logarithmic plots on the right column give a clearer insight on the modulation properties. The I-V characteristics are modulated in a voltage range of $\Delta_w/e$ corresponding to a swing in current of few order of magnitude that can further increase by lowering $\Gamma$.

Considering an electrical setup where the $\omega$-SQUIPT is biased with a proper current $I_b$, the voltage drop depends on flux, giving the flux to voltage characteristics  $V(\Phi)$ (see Fig.  \ref{fig:sketch}).
The optimal voltage-gap swing $\Delta_w/e$ can be approached at low current bias, making the $\omega$-SQUIPT a low power dissipation magnetometer.
In figure \ref{fig:F2V_SCP} on the left side the flux to voltage characteristics $V(\Phi)$ is plotted for two representative devices with $E_{Th}/\Delta_0=0.5$ and $E_{Th}/\Delta_0=5$ in the case of both NP and SCP. The main modulation interval is $\Phi/\Phi_0=[-1/3,1/3]$ (with $\Phi_0$ periodicity); on the contrary, the interval $\Phi/\Phi_0=[2/3,4/3]$ has a flat trend.
The performance of the device as a magnetometer can be estimated by the flux to voltage transfer function
\begin{equation}
\mathcal{F}(\Phi) = \frac{\partial V_I}{\partial \Phi} \, \, .
\end{equation}
The $\mathcal{F}$ function is reported  in Fig. \ref{fig:F2V_SCP} on the right side. The performance in terms of magnetometry of the $\omega$-SQUIPT is 3.8$\Delta_0/e\Phi_{0}$ for $E_{Th}/\Delta_0=0.5$ and 4.3$\Delta_0/e\Phi_{0}$ for $E_{Th}/\Delta_0=5$. 

We note that these performances are lower than those of a conventional SQUIPT. Indeed, for sake of comparison it is sufficient to consider the total flux on the device. The total flux interval of the main modulation is from zero to the closure of the induced minigap. In the SQUIPT, the minigap closes at $\Phi_0/2$; in a $\omega$-SQUIPT, the gap closes at total flux $2\Phi_0/3$, that is greater than the SQUIPT case. This means that a certain swing in the output signal requires a greater flux variation in the $\omega$-SQUIPT, lowering then its sensitivity.

Nevertheless, the $\omega$-SQUIPT has also interesting gradiometric properties. Let us consider the region around the fluxes point $(\Phi_L/\Phi_0,\Phi_R/\Phi_0)=(1/2,1/2)$ in the $N_F$ plot for full symmetric $\omega$-SQUIPT (Fig.  \ref{fig:dyn_VS_in}). Along the diagonal line $\Phi_L=\Phi_R$, the modulation is smaller with respect to other directions. In particular, it reaches the maximum value for $\Phi_L=-\Phi_R$. Hence, the sensitivity is greater for magnetic fields with a spatial gradient. Gradiometric properties are exploited for magnetic measurement protected from noise caused by far source\cite{clarke2006}.
\begin{figure}[htp]
\centering
\includegraphics[width=8.5cm]{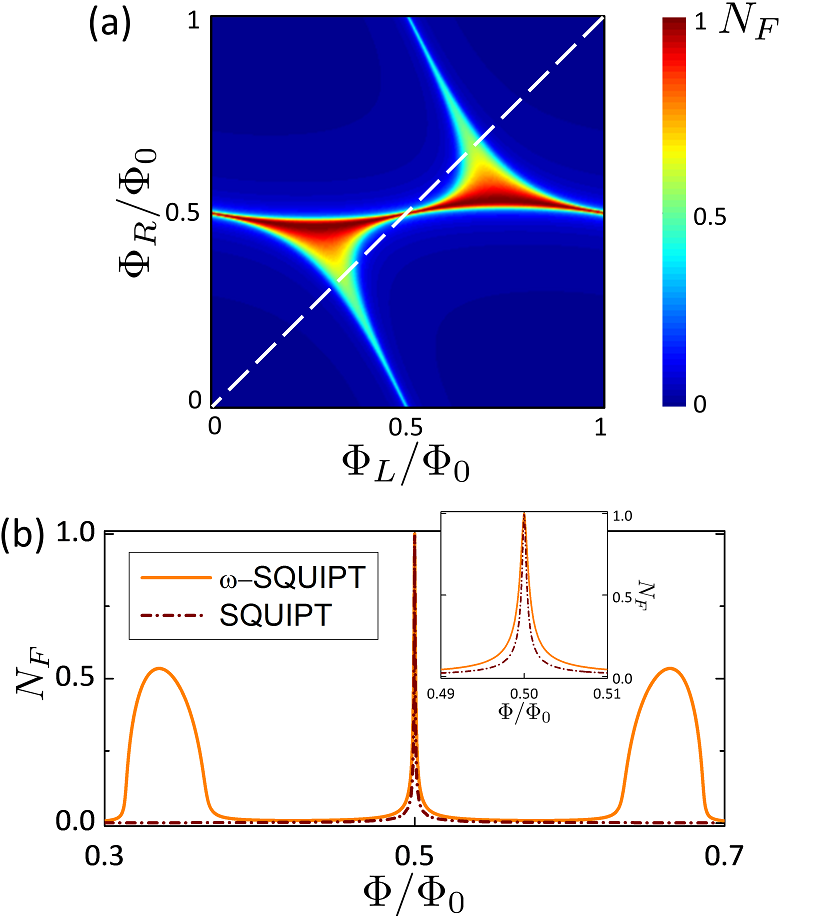}
\caption{(Color online) Working principle of the $\omega$-SQUIPT as a magnetometer. Here the interfaces are asymmetric, with  $r_R=r_C=0.1$ and $r_L=5$.
Panel (a) presents the DoS at fermi energy $N_F$ as a function of $\Phi_L$ and $\Phi_R$.
Panel (b) DoS of the $\omega$-SQUIPT  at Fermi energy $N_F$ in the case of equal fluxes $\Phi_L=\Phi_R=\Phi$ (orange solid line). For sake of comparison we plot also the result in the case of a conventional two-terminal device (dark red dashed curve) with equal contact resistances $r=0.1$.}
\label{fig:skew_mag}
\end{figure}

Finally, we comment on a different possible application of the $\omega$-SQUIPT as a magnetometer. Basically, this possibility relies on the shape of the DoS at Fermi energy of the three-terminal JJ. Consider, for example, an $\omega$-SQUIPT whose S/N resistances are asymmetric with $r_R=r_C=0.1$ and $r_L=5$, as in Fig. \ref{fig:skew_mag}.
In this case, the shape of the  DoS at Fermi energy $N_F$ is skewed (Fig. \ref{fig:skew_mag}). A symmetric flux that goes from $\Phi=0$ to $\Phi=\Phi_0$ (Fig. \ref{fig:skew_mag}, white line in panel (a)) crosses the red conductive region in three different points. In these crossing points, the DoS at Fermi energy shows peaks depending on the flux $\Phi$. 
The strong modulation of $N_F$ can be exploited for magnetometry. Notice that here, the experimental setup should be different from the current biased setup discussed above. For example, a lock-in configuration that measures the differential conductance at zero voltage can be used. Panel (b) of Fig.\ref{fig:skew_mag} shows the cuts of the DoS at Fermi energy for equal fluxes in the asymmetric configuration with $r_R=r_C=0.1$ and $r_L=5$ (orange solid line). For sake of comparison we have also plotted the analogous quantity for a conventional SQUIPT\cite{dambrosio_normal_2015} with opacities $r=0.1$ (dark red dashed curve).
As one can argue from the figure, also the two-terminal device can be used as a magnetometer, since it presents a peaked structure around $\Phi=\Phi_0/2$ \citep{dambrosio_normal_2015}. The inset depicts a magnification in the region near $\Phi=\Phi_0/2$, showing that the peak is sharper in the case of a conventional SQUIPT.
Nevertheless, the $\omega$-SQUIPT has also other intervals of modulation in $\Phi=(0.34\pm 0.03)\Phi_0$ and $\Phi=(0.66\pm 0.03)\Phi_0$, demonstrating that it has a larger region of working points as a magnetometer.

\section{Conclusions}
\label{sec:conclusion}
In summary, the paper reports an exaustive theoretical investigation of different coherent transport properties of a three-terminal hybrid device, the so-called $\omega$-SQUIPT.
By means of a full numerical solution of the Usadel equation, extended to the case of three S leads, we have studied the effects on the proximized metallic nanowire of the length, the inelastic scattering and the quality of the S/N interfaces.
 We have shown that the spectral properties are an useful tool to identify transitions between gapless and gapped states in this three-terminal setup.
 The induced supercurrents in the different arms of the device are discussed in detail, showing that these can be an alternative allmark of non-trivial topological properties.
The quasi-particle transport properties through a metallic probe tunnel-coupled to the Josephson junction are presented both in the case of a metallic and a superconducting probe.
Since the $\omega$-SQUIPT is sensitive to magnetic fluxes, we have inspected its magnetometric features, finding that this device can have potential applications as a gradiometer or magnetometer.
Finally, we emphasize that the theoretical results reported here can serve as a starting point for a better fundamental understanding of multi-terminal JJs which recently have drawn great interest due to their exotic properties and potential applications in quantum computing.
\begin{acknowledgments}
\label{sec:Acknowledgements}

F.V. and F.G. acknowledge the European Research Council  under  the  European  Union’s  Seventh  Framework  Program  (FP7/2007-2013)/ERC  Grant  Agreement  No.  615187-
COMANCHE  and  MIUR-FIRB2013—Project  Coca  (Grant No.  RBFR1379UX)  for  partial  financial  support.  The  work of E.S. was funded by a Marie Curie Individual Fellowship (MSCA-IFEF-ST No. 660532-SuperMag). The work of F.S.B. was supported by Spanish Ministerio de Economia y Competitividad (MINECO) through Project No. FIS2014-55987-P.

\end{acknowledgments}

\end{document}